\documentclass{article}
\usepackage{spconf,amsmath,graphicx}
\usepackage{caption}
\usepackage{amsfonts}
\usepackage{booktabs}
\usepackage{multirow}
\usepackage{color}
\usepackage{subcaption}
\usepackage{float}
\usepackage{color}
\usepackage{amsthm,amsmath,amssymb}
\usepackage{mathrsfs}
\usepackage{bm}

\title{LOW-DIMENSIONAL DENOISING EMBEDDING TRANSFORMER FOR ECG CLASSIFICATION}
\name{Jian Guan$^1$, Wenbo Wang$^1$,  Pengming Feng$^2$, Xinxin Wang$^3$, and Wenwu Wang$^4$\thanks{This work was supported by the Natural Science Foundation of Heilongjiang Province under Grant No. YQ2020F010.}}
\address{
  $^1$Group of Intelligent Signal Processing, Harbin Engineering University, Harbin, 150001, China\\
  $^2$State Key Laboratory of Space-Ground Integrated Information Technology, Beijing, 100095, China\\
  $^3$Alibaba Group, Hangzhou, 310051, China\\
  $^4$Centre for Vision Speech and Signal Processing, University of Surrey, Guildford, GU2 7XH, UK}
\begin{document}
%
\maketitle
\begin{abstract}
\label{abs}
The transformer based model (e.g., FusingTF) has been employed recently for Electrocardiogram (ECG) signal classification. However, the high-dimensional embedding obtained via 1-D convolution and positional encoding can lead to the loss of the signal's own temporal information and a large amount of training parameters.  
 In this paper, we propose a new method for ECG classification, called low-dimensional denoising embedding transformer (LDTF), which contains two components, i.e., low-dimensional denoising embedding (LDE) and transformer learning. In the LDE component,  a low-dimensional representation of the signal is obtained in the time-frequency domain while preserving its own temporal information. 
And with the low-dimensional embedding, the transformer learning is then used to obtain a deeper and narrower structure with fewer training parameters than that of the FusingTF. Experiments conducted on the MIT-BIH dataset demonstrates the effectiveness and the superior performance of our proposed method, as compared with state-of-the-art methods.
\end{abstract}
\section{Introduction}
\label{sec:intro}
\begin{figure*}[ht]	
	\centering
	\setlength{\belowcaptionskip}{1pt}
	\setlength{\abovecaptionskip}{1pt}
	\includegraphics[width=.85\textwidth]{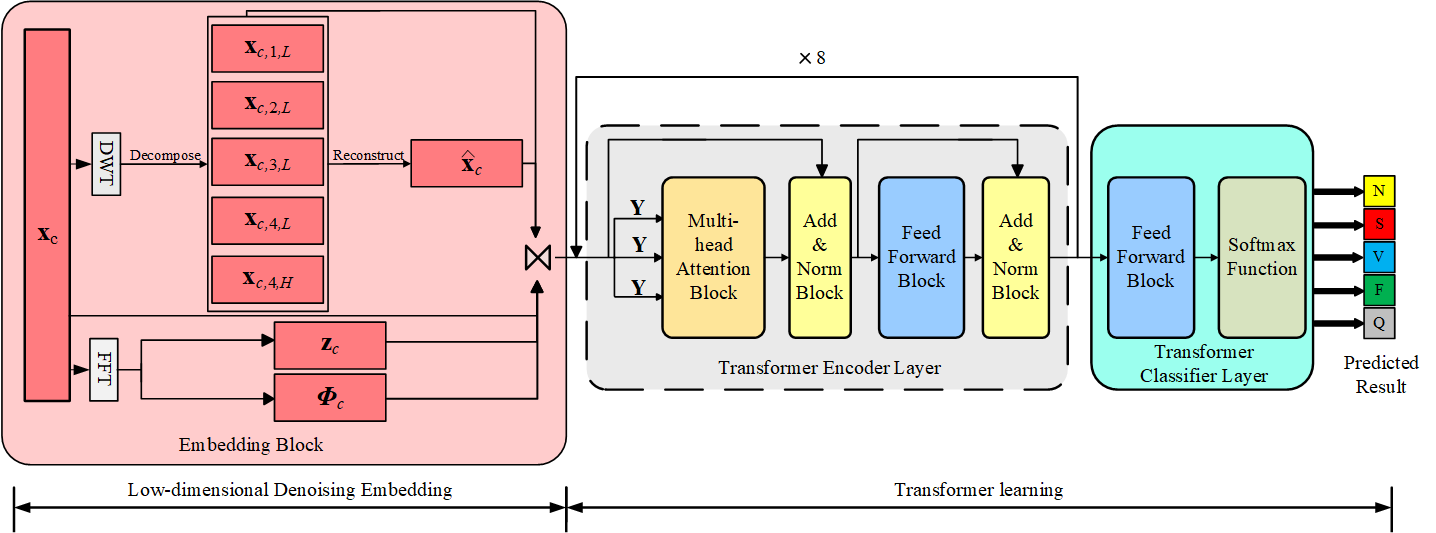}
	\caption{Framework of the proposed LDTF model, which consists of two stages: low-dimensional denoising embedding  and transformer learning. Here ``$\Join$'' denotes the concatenation operation, and $c$ denotes the $c$-th channel. }
\label{fig:1}
\end{figure*}
Electrocardiogram (ECG) is widely used in clinical diagnosis to provide information about cardiac disease according to its three different components: P wave, QRS complex, and T wave in different conditions \cite{zimetbaum2003use}. The characteristics of these waveforms are of great importance for the classification of the ECG signals and the diagnosis of cardiovascular disease.

Deep learning based models \cite{acharya2017deep, el2020ecg, li2018deep, al2019classification, yildirim2018novel, lynn2019deep} have been used for ECG classification, giving state-of-the-art performance. According to how the information is exploited, these approaches can be divided into two categories: methods based on convolutional neural network (CNN) and methods based on sequence module. 
The CNN-based methods aim to exploit morphological features of ECG signals either in one dimensional time series \cite{acharya2017deep, el2020ecg} or in two dimensional image \cite{li2018deep}, and are shown to outperform the traditional machine learning methods such as support vector machines (SVM) \cite{abdalla2019ecg, el2019efficient, zhang2019comparing, ge2019ecg}.  However, these methods focus on the morphological features in the area centred around a certain peak data point or in the wavelets decomposed from original signal via DWT \cite{el2020ecg}, and they are ineffective in exploiting the latent feature of the temporal signal in either time or frequency domain.   
 In contrast, the methods with sequence module structure often employ recurrent neural network (RNN) based models, such as long-short term memory (LSTM) \cite{al2019classification, yildirim2018novel} or gated recurrent unit (GRU) \cite{lynn2019deep}, to capture the latent temporal information of the ECG signals for classification.

Recently, the transformer model has been proposed and applied to many applications including ECG classification \cite{ yan2019fusing}, giving state-of-the-art performance \cite{vaswani2017attention}. In \cite{ yan2019fusing}, a transformer based method, i.e., FusingTF, is presented for ECG classification, where the ECG signal is represented in a high-dimensional space (i.e., 64-D) via a simple 1-D convolutional layer to expand the original dimension (i.e., 1-D) for the transformer. After this, a positional encoder is applied to incorporate additional temporal information into the 64-D result,  however, this method is limited in compensating for the loss of temporal information within the signal caused by the 1-D CNN.  
In addition, the high-dimensional embedding will result in a model with increased number of parameters. Although the original ECG signal could be used directly as the transformer input, the presence of noise in the signal can degrade the classification performance. Representing the ECG signal in the embedding space can help denoise the signal and reduce the number of parameters in the model. 

In this paper, we present a low-dimensional denoising embedding transformer model (LDTF) for ECG classification, which improves the performance with fewer training parameters by introducing a low-dimensional denoising embedding (LDE) approach for signal embedding and a deeper and narrower transformer model for signal classification.  
The proposed method includes two stages, namely, low-dimensional denoising embedding and transformer learning.  First, the original ECG signal is represented in a low-dimensional space by the LDE approach, which employs discrete wavelet transform (DWT) \cite{yildirim2018novel, martis2013ecg, sahoo2017multiresolution} and fast Fourier transform (FFT) to exploit the features from both time domain and frequency domain simultaneously. 

The DWT of ECG signal can effectively expand the dimension of the original ECG signal and extract the primary features of the signal in time domain thereby preserving the temporal information of the signal, whereas FFT generates the features of the ECG signal in the frequency domain.  Instead of mixing  different kinds of features obtained above, we integrate them together by representing each feature with a specific dimension in the low-dimensional space. 
Such that, our LDE method can well preserve the characteristics of ECG signals. 
Then, with the low-dimensional embedding results for transformer learning,  we can thereby use a narrower and deeper structure by stacking more transformer encoder layers to improve the classification performance with fewer training parameters in each layer. The experiments conducted on the MIT-BIH dataset  \cite{932724} verify the effectiveness of our proposed model.

The remainder of the paper is organized as follows: Section 2 presents the proposed model in details; Section 3 shows the experimental results; and Section 4 summarizes the paper and gives the conclusion.

\section{Proposed Method}
\label{sec:method}
In this section, we  introduce the proposed LDTF model for ECG classification, which consists of a low-dimensional denioising embedding stage and a transformer learning stage. 
The framework of the proposed model is given in Figure \ref{fig:1}. The details of the model are given next.
\subsection{Low-Dimensional Denoising Embedding}
\label{sec:embedding}
In this stage, we propose an LDE method to well represent the original ECG signal in a low-dimensional space, where DWT and FFT are employed to extract the features from both time domain and frequency domain.

Here, DWT is used to decompose and reconstruct the original ECG signal to get its  waveform representations in time domain, which aims to preserve the signal's own temporal information. For the purpose of a more comprehensive representation of the signal, fast Fourier transform is also adopted to extract features of the signal (i.e., magnitude and phase angle)  in  frequency domain. Then, we can obtain a low-dimensional embedding representation for transformer learning by concatenating the DWT decomposed wavelets, reconstructed signal, the original ECG signal and its magnitude and phase angle. As this representation already contains the information from both time and frequency domain, there is no need to use other function as in \cite{yan2019fusing} to add additional temporal information to the embedding result. The details of the LDE method is given as follows.

Let $\mathbf{X} \in {\mathbb{R}}^{2 \times \ell}$ be the two leads ECG signal and $\mathbf{x}_c \in {\mathbb{R}}^{1 \times \ell}$ be one lead of the ECG signal, where $c \in \{1, 2\}$ is the channel number and $\ell$ is the length of the signal. 
Then, it can be decomposed to wavelets via DWT, as follows  
%
\begin{equation}
\label{eq:1}
{\mathbf{x}_{c,j,\mathit{H}}}[n] = \sum^{K-1}_{k=0} {{\mathbf{x}_{c,j-1,\mathit{H}}}[2n-k]\mathbf{h}(k)}
\end{equation}
\begin{equation}
\label{eq:2}
{\mathbf{x}_{c,j,\mathit{L}}}[n] = \sum^{K-1}_{k=0} {{\mathbf{x}_{c,j-1,\mathit{L}}}[2n-k]\mathbf{g}(k)}
\end{equation}
where $n$ denotes the $n$-th element of the signal and $K$ is the length of slide window, $k \in \{0, \cdots, K-1\}$. Here, $\mathbf{h}(\cdot)$ and $\mathbf{g}(\cdot)$ denote the high pass filter and low pass filter, respectively. ${\mathbf{x}_{c,j,\mathit{H}}}$ and ${\mathbf{x}_{c,j,\mathit{L}}}$ are the decomposing results at $j$-th level of $\mathbf{h}(\cdot)$ and $\mathbf{g}(\cdot)$ respectively,  where $j \in \{1,2,3,4\}$.
 Then, we can obtain the denoised reconstruction signal $\mathbf{\hat{x}}_c$ with an inverse decomposition operation as follows 
\begin{equation}
\label{eq:3}	
{\mathbf{\hat{x}}}_{c,j-1,H} = \sum^{K-1}_{k=0} {\mathbf{\hat{x}}_{c,j,\mathit{H}}}[2n-k]\mathbf{h}(k) + \sum^{K-1}_{k=0}{ {\mathbf{x}_{c,j,\mathit{L}}}[2n-k] \mathbf{g}(k)}
\end{equation}
when $j=1$, we will obtain the reconstructed signal  $\mathbf{\hat{x}}_c = {\mathbf{\hat{x}}}_{c,0,H}$.

To obtain a better representation of the signal, we also extract the information from frequency domain, where the magnitude $\mathbf{z}_c \in \mathbb{R}^{1 \times \ell}$ and the phase angle $\bm{\phi}_c \in \mathbb{R}^{1 \times \ell}$ of the signal $\mathbf{x}_c$ are used as parts of the embedding result, which are obtained respectively, as follows
\begin{equation}
\label{eq:4}	
\mathbf{z}_c = |fft(\mathbf{x}_c)|
\end{equation}
\begin{equation}
\label{eq:5}	
\bm{\phi}_c = angle(fft(\mathbf{x}_c)) 
\end{equation}
where $fft(\cdot)$ denotes the fast Fourier transform operation. Here, $|\cdot|$ and $angle(\cdot)$ denote the calculation of absolute value and phase angle respectively. 

Finally, we can obtain a low-dimensional embedding result $\mathbf{X}_{lde} \in \mathbb{R}^{d \times \ell}$ by concatenating the two leads original ECG signals, decomposed wavelets, reconstructed signals, and magnitudes and phase angles, which is expressed as follows
\begin{equation}
\label{eq:6}
    \begin{aligned}	
		\mathbf{X}_{lde} = [ & \mathbf{x}_1; {\mathbf{x}_{1,1,\mathit{L}}}; \cdots;{\mathbf{x}_{1,4,\mathit{L}}}; {\mathbf{x}_{1,4,\mathit{H}}};\mathbf{\hat{x}}_1;\mathbf{z}_1; \bm{\phi}_1;\\
		 & \mathbf{x}_2; {\mathbf{x}_{2,1,\mathit{L}}}; \cdots;{\mathbf{x}_{2,4,\mathit{L}}}; {\mathbf{x}_{2,4,\mathit{H}}};\mathbf{\hat{x}}_2;\mathbf{z}_2; \bm{\phi}_2] 
     \end{aligned}
\end{equation}
Here the dimension $d$ of $\mathbf{X}_{lde}$ is 18.  Because of integrating different features in different dimensions separately, the latent features of $\mathbf{X}_{lde}$ can be more easily learned by the transformer model.

\subsection{Transformer Learning}
\label{ssec:classification}
Regarding transformer learning stage, thanks to the low dimensional embedding result as the input, we can obtain a narrower and deeper structure with less training parameters for ECG classification. This is because that here the space consuming  mainly comes from the fully forward connected network structure (i.e., feed forward block) in each transformer encoder layer.
Therefore, with the low-dimensional input, the training parameters of each transformer encoder layer will be reduced rapidly. Such that, we can stack more transformer encoder layers with fewer parameters (i.e., 9,258,742) in each layer, and fewer parameters (i.e., 74,087,228) for the whole model. Whereas the similar study in \cite{yan2019fusing}, i.e., FusingTF, has 32,029,382 parameters in each transformer encoder layer, and a total number of 192,237,988 parameters.
 In addition, this deeper and narrower structure can effectively improve the classification performance, which will be demonstrated in Section \ref{sec:3_2}. 

The transformer learning part of our model consists of 8 transformer encoder layers and one transformer classifier layer. Each transformer encoder layer includes one multi-head attention block (MAB) with 6 heads, one feed forward block (FFB) and two add \& norm blocks (ANB) as shown in Figure \ref{fig:1}. 

Assuming $\mathbf{Y}$ is the input of MAB in each transformer encoder layer, which is initially set as $\mathbf{Y}=\mathbf{X}_{lde}$. Then, the linear mapping result $\mathbf{Y}\mathbf{W}_{i,q}$ of $\mathbf{Y}$ via transform matrix $\mathbf{W}_{i,q}\in \mathbb{R}^{\ell \times \ell}$ can be used to calculate the attention result $\bm{\mathcal{H}}_i \in \mathbb{R}^{d \times \ell}$ of the $i$-th head, where $i \in \{1, \cdots, 6\}$ and $q \in \{1, 2, 3\}$. Such that $ \bm{\mathcal{H}}_i$ can be obtained as follows 
\begin{equation}
\label{eq:7}
	\bm{\mathcal{H}}_i = {\rm{softmax}}(\frac{(\mathbf{Y}\mathbf{W}_{i, 1})(\mathbf{Y}\mathbf{W}_{i, 2})^{\top}}{\sqrt{d}})(\mathbf{Y}\mathbf{W}_{i, 3})
\end{equation}
where $\top$ denotes the transposition operation, and the softmax function is used to calculate the weights on the linear mapping result $\mathbf{Y}\mathbf{W}_{i, 3}$.

To integrate the attention results of all heads, here we perform another linear mapping operation via transform matrix $\mathbf{W}_o \in \mathbb{R}^{6 \ell \times \ell}$, then the output of MAB can be obtained as follows

\begin{equation}
\label{eq:8}
\bm{\mathcal{O}}_{MAB} = [\bm{\mathcal{H}}_1, \cdots,\bm{\mathcal{H}}_i \cdots, \bm{\mathcal{H}}_6]\mathbf{W}_o
\end{equation}

After that, a residual connection and layer norm operation are conducted in ANB to connect the FFB, here the FFB consists of two fully connected layers with a rectified linear unit (ReLU).  Then, with an ANB again after the FFB, we can obtain the output of the transformer encoder layer, which will be used as the input of the next transformer encoder layer or the transformer classifier layer  that contains one fully connected layer and a softmax function. 
Finally, we will get the ECG classification results after the transformer learning.

\section{Experiments and Results}
\label{sec:expriments}
\subsection{Experimental Setup}
\label{ssec:dataset}
\textit{\textbf{Dataset}} 
We evaluate the proposed model on MIT-BIH, 
which are the two-channel recording ECG signals sampled from 48 different patients at a 360Hz sampling frequency for 30 minutes with 23 types. In our work, 5 essential arrhythmia groups from MIT-BIH are employed for evaluation, which are described in Table \ref{Tab:1}. These heartbeats classes are categorized by the American Association of Medical Instrumentation (AAMI). Here, 112,551 ECG segments are selected for experimental evaluation, where each segment is sampled with 241 sampling points centred at R peak point, which means the length of the signal is set as $\ell = 241$ in our experiments. 
\begin{table}[ht]
\small
	\setlength{\belowcaptionskip}{1pt}
	\setlength{\abovecaptionskip}{1pt}
	\centering
	\caption{ ECG signal classification standard specified by ANSI/AAMI EC57 and the number of each class in our dataset}
	\label{Tab:1}
	\begin{tabular}{lll}
		\toprule
Groups & ECG classes & Number\\

\midrule
N               & Normal(N)                                    &75,017\\                      
				& Left Bundle Brunch Block(L)                  &8,071\\
				& Right Bundle Brunch Block(R)                 &7,255\\
				& Atrial Escape (e)                            &16\\
				& Nodal (junctional) escape(j)                 &229\\ 
\cline{1-3}
S               & Atrial Premature(A)                          &2,546\\
				& Aberrant Atrial Premature(a)                 &150\\
				& Nodal (Junctional) Premature(J)              &83\\
				& Supra-ventricular Premature(S)               &2\\
\cline{1-3}
V               & Premature Ventricular Contraction(V)         &7,129\\
				& Ventricular escape(E)                        &106\\ 
\cline{1-3}
F               & Fusion of Ventricular and Normal(F)          &802\\ 
\cline{1-3}
Q               & Paced(/)                                     &7,023\\
				& Fusion of Paced and Normal(f)                &982\\
				& Unclassifiable(Q)                            &33\\
\bottomrule
\end{tabular}
\end{table}
\begin{table}[ht]
\small
\centering
	\setlength{\belowcaptionskip}{1pt}
	\setlength{\abovecaptionskip}{1pt}
\caption{Hyperparameters settings for model training}
\label{Tab:02}
	\resizebox{0.48\textwidth}{10.5mm}{
\begin{tabular}{c}
\toprule
Selected hyperparameters\\
\midrule
Loss function = Cross-Entropy, Optimizer = SGD,\\ Learning rate = 0.001, Batchsize = 64, Dropout ratio = 10\%\\
ECG segment length = 241, Head number = 6\\
\cline{1-1}
\bottomrule
\end{tabular}}
\end{table}
\begin{table}[ht]
\small
	\setlength{\belowcaptionskip}{1pt}
	\setlength{\abovecaptionskip}{1pt}
\centering
\caption{Validation of the LDE method and the deeper and narrower structure of our model}
\label{Tab:03}
	\resizebox{0.48\textwidth}{13.5mm}{
\begin{tabular}{ccccccc}
\toprule
Category & \multicolumn{3}{c}{Rec(\%)} & \multicolumn{3}{c}{Pre(\%)} \\
\cmidrule(r){2-4} \cmidrule(r){5-7}
&  LDTF      &  FTFD  	&  FusingTF
&  LDTF      &  FTFD   	&  FusingTF
\\
\midrule
 N 	&\textbf{95.77}	   & 94.40    & 93.20               		&\textbf{96.81}    & 93.60    & 96.30       \\
 S 	&\textbf{97.65}	   & 96.20    & 93.39                   		&\textbf{97.58}    & 96.87    & 96.72     		\\
 V 	&\textbf{97.72}	   & 97.03    & 93.45              		&\textbf{97.38}    & 95.51    & 97.43          		\\
 F 	&\textbf{99.01}	   & 95.83    & 95.12                   		&\textbf{98.38}    & 97.55    & 96.33          		\\
\cline{1-7}
 Average 	&\textbf{97.54}	   & 95.87    & 93.79                   		&\textbf{97.54}    & 95.88    & 96.70          	\\
\cline{1-7}
\bottomrule
\end{tabular}}
\end{table}

To alleviate the categories unbalanced problem and remove the amplitude scaling problem, synthetic minority oversampling technique (SMOTE) and Z-score normalization are adopted for data preprocessing respectively \cite{pandey2019automatic}. Then, we select $80\%$ ECG segments from each class as training set, and the remaining $20\%$ ECG segments from each class as test set. The number of each category is also given in Table \ref{Tab:1}.  The  hyperparameters for our model training is given in Table \ref{Tab:02}.

\noindent\textit{\textbf{Performance Metrics}} We use the performance metrics in terms of Recall (Rec) and Precision (Pre) to evaluate the ECG classification performance of the proposed model,  as follows
\begin{equation}
\label{eq:9}
\rm Rec = \frac{TP}{TP + FN}
\end{equation}
\begin{equation}
\label{eq:10}
\rm Pre = \frac{TP}{TP + FP}
\end{equation}
where TP and TN denote the  true positive and the true negative respectively. FP and FN denote  the false positive and  the false negative respectively.
\subsection{Experimental Results}
\label{sec:3_2}
\textit{\textbf{Ablation Study}}
 To show the effectiveness of both LDE and the deeper and narrower structure, 
 an ablation study is conducted for a four categories ECG classification task, which is the same as FusingTF in \cite{yan2019fusing} for the purpose of fair comparison. The baseline is FusingTF with 6 transformer encoder layers, and its embedding stage consists of a 1-D CNN layer to embed the original ECG signal into 64-D space and a positional encoding operation via a sinusoidal function. 
 
Here, we simply modify FusingTF with 8 transformer encoder layers, and use the 1-D CNN layer to embed the signal into an 18-D space, such that we can obtain a deeper and narrower structure, i.e., FusingTF-Deep (FTFD). Compared to LDTF, the only difference is that FTFD used same embedding method (i.e., 1-D CNN and a positional encoder) as FusingTF for embedding. The experimental results are given in Table \ref{Tab:03}.
From Table \ref{Tab:03}, we can see that FTFD outperforms FusingTF, which demonstrates the validity of the deeper and narrower structure. The proposed LDTF can achieve the best performance, as compared with FTFD and the baseline FusingTF, which also verifies the effectiveness of the LDE method.

\begin{table}[ht]
\small
	\setlength{\belowcaptionskip}{1pt}
	\setlength{\abovecaptionskip}{1pt}
\centering
\caption{Performance comparison in terms of average Recall and average Precision}
\label{Tab:04}
	\resizebox{0.45\textwidth}{9mm}{
\begin{tabular}{ccccc}
\toprule
Method & \multicolumn{2}{c}{Average Recall(\%)} & \multicolumn{2}{c}{Average Precision(\%)}\\
\cline{1-5}
\textbf{LDTF} & \multicolumn{2}{c}{\textbf{98.39}}& \multicolumn{2}{c}{98.41}\\
WCNN & \multicolumn{2}{c}{93.55}& \multicolumn{2}{c}{96.72}\\
MWFE & \multicolumn{2}{c}{96.86}& \multicolumn{2}{c}{\textbf{98.92}}\\
Ad-CNN & \multicolumn{2}{c}{93.90}& \multicolumn{2}{c}{98.90}\\
\cline{1-5}
\bottomrule
\end{tabular}
}
\end{table}

\noindent\textit{\textbf{Performance Comparison}} We also compare our proposed method with other state-of-the-art methods (i.e., MWFE \cite{sahoo2017multiresolution}, Ad-CNN \cite{kiranyaz2015real} and WCNN \cite{el2020ecg}) for five categories ECG classification task. All these three methods use DWT for data preprocessing. MWFE and Ad-CNN only apply DWT for denoising, whereas WCNN uses the decomposed wavelets obtained by DWT for feature learning. The results are given in Table \ref{Tab:04}. From Table \ref{Tab:04}, we can see our proposed LDTF can achieve the best performance in terms of average Recall, 
 as compared with other state-of-the-art methods. Although, the average precision of LDTF is slightly lower than MWFE,  LDTF improves the overall performance.

\section{Conclusion}
\label{sec:concl}
In this paper, we presented a low-dimensional denoising embedding transformer model for ECG classification, where a low-dimensional embedding method is proposed for the well representation and embedding the signal into low-dimensional space for transformer learning. The proposed method can not only significantly reduce the training parameters, but also improve the overall performance for ECG classification. Experimental results demonstrated the effectiveness and the superiority of the our method, as compared with other state-of-the-art methods. 
%

\bibliographystyle{IEEEtran}
\bibliography{refs}

\end{document}